\documentclass[aps,prl,twocolumn,floatfix,floats,showpacs,superscriptaddress,raggedbottom]{revtex4}
\usepackage{amsmath,amsfonts,amssymb,amsthm}
\usepackage{color}
\usepackage{graphicx,latexsym}
\usepackage{dcolumn}

\begin{document}

\title{Coherent Nonlinear Quantum Model for Composite Fermions}
\author{Gilbert Reinisch}
% \affiliation{Universit\'e de Nice - Sophia Antipolis, CNRS, Observatoire de la C\^ote d'Azur,
 %            BP 4229, 06304 - Nice C\'edex 4, France}
\author{Vidar Gudmundsson}
\affiliation{Science Institute, University of Iceland, Dunhaga 3, IS-107 Reykjavik,
             Iceland}
\author{Andrei Manolescu}
\affiliation{School of Science and Engineering, Reykjavik University, Menntavegur 1, IS-101 Reykjavik,
             Iceland}

\begin{abstract}
Originally  proposed by  Read 
\cite{Read89:62}
and Jain \cite{Jain89:199}, the so-called ``composite-fermion'' is a phenomenological attachment of 
two infinitely thin local flux quanta seen as nonlocal vortices to two-dimensional (2D) electrons 
embedded in a strong orthogonal magnetic field. In this letter, it 
is described as a highly-nonlinear and coherent mean-field  quantum process of the soliton type by use of a  
2D stationary Schr{\"o}dinger-Poisson differential model with only two Coulomb-interacting electrons. 
At filling factor $\nu=\frac{1}{3}$ of the lowest Landau level,
it agrees with both the exact two-electron antisymmetric Schr{\"o}dinger wave function
and Laughlin's Jastrow-type guess for the fractional quantum Hall effect, hence providing  
this later with a tentative physical justification based on first principles.
\end{abstract}


\maketitle
%
% \section{Introduction}
%
Perhaps the most spectacular physical concept introduced in the  description of 
Fractional Quantum Hall Effect (FQHE) is Composite Fermion (CF). It consists in an
intricate mixture of $N_e$ electrons and vortices in a two-dimensional (2D) electron 
gas orthogonal to a (strong) magnetic field
such that the lowest Landau level (LLL) is only partially occupied.
Actually, the CF concept  provides an intuitive phenomenological way of looking at 
electron-electron correlations as a part of
sophistiscated many-particle quantum effects where charged electrons do avoid 
each other by correlating
their relative motion in  the energetically most advantageous fashion conditioned by the magnetic field.
Therefore it is picturesquely assumed that each electron lies at the center of a 
vortex whose trough 
represents  the outward displacement of all fellow electrons and, hence,
accounts for actual decrease of their mutual repulsion \cite{Read89:62,Stormer99:875}. 
Or equivalently, in the simplest case of
$N_e=2$ electrons  considered in the present letter, that two
flux quanta $\Phi_0=hc/e$ are {\it ``attached''} to each electron, turning
the pair into a LLL of two CFs with a $6\Phi_0$ resulting flux \cite{Jain89:199}. 
The corresponding Aharonov-Bohm quantum phase shift equals  $2\pi$. In addition to the $\pi$ 
phase shift of core electrons, it agrees with the requirements
of the Laughlin correlations expressed by the Jastrow polynomial of degree $3$
and corresponding to the LLL filling factor $\nu = \frac{1}{3}$ \cite{Laughlin83:1395,Laughlin83:3383}.
Laughlin's guessed wavefunction for odd  polynomial  degree was  soon  
regarded as a Bose condensate \cite{Girvin87:1252,Zhang89:62,Rajaraman96:793}
whereas for even degree, it was considered  as
a mathematical artefact describing a Hall metal
that consists of  a well defined Fermi surface
at a vanishing magnetic field generated by a Chern-Simons
gauge transformation of the state at exactly
$\nu = \frac{1}{2}$ \cite{Halperin93:47,Rezayi94:72}.

Although they provide a simple appealing single-particle illustration
of Laughlin correlations, the physical origin of the CF auxiliary field fluxes
remains  unclear. In particular, the way they are fixed to particles is not
explained. Hence tentative theories avoiding the CF concept like e.g.\ a recent
topological formulation of FQHE \cite{Jacak12:26}. 
In the present letter, we show how a strongly-nonlinear mean-field quantum model  
provides an alternative Hamiltonian physical description, based on first principles,
of the debated CF quasiparticle.

Consider the 2D electron pair confined in the $x-y$
plane under the action of the orthogonal magnetic field $ {\bold B} $. It
is situated at $z_{1,2}=x_{1,2}+iy_{1,2}$.
Adopt the usual center-of-mass ${\bar z}=(z_1+z_2)/2$ and internal coordinate 
$z =(z_1-z_2)/\sqrt{2}$ separation and select odd-$m$
angular momenta $m\hbar$ in order to comply with the antisymmetry of the two-electron  wavefunction 
under electron interchange. The corresponding  internal motion radial eigenstate 
$\Psi_m(x,y)=u_m(r)e^{im\phi}$  with $z =x +iy =re^{i\phi}$
is defined in units of length and energy
by the cyclotron length $\lambda_c=\sqrt{\hbar / (M\omega_c)}$  
and by the Larmor energy $\hbar \omega_L=\frac{1}{2} \hbar \omega_c = \hbar e B/(2Mc)$
where $M$ denotes the effective mass of the electron which may
incorporate many-body effects. The eigenstate $u_m$ is given by  \cite{Laughlin83:3383}:
\begin{equation}
\label{eq-SchroeRBL}
      \Bigl[\nabla_X^2   +   E_m +m 
      -\frac{m^2}{X^2} -\frac{X^2}{4} - \frac{K}{X} \Bigr] u_m =0.
\end{equation}
The radial part of the 2D Laplacian operator is $\nabla_X^2=d^2/dX^2 + X^{-1}(d/dX)$, the
energy eigenvalue is  $E_m$
and $X=r/\lambda_c$. The dimensionless parameter
\begin{equation}
      \label{eq-K}
      K=\sqrt{2}\,\,\frac{ e^2/(\epsilon \lambda_c )}{\hbar\omega_c}
,
\end{equation}
where $\epsilon $ is the dielectric constant of the semiconductor host,
compares  the Coulomb interaction between the two particles with the cyclotron energy. Obviously,
$K=0$ corresponds to the free-particle case.
Actually, the internal motion could  be approximated by the 2D free-particle 
harmonic oscillator eigenstate $|m\rangle$ as long as $K\leq \sqrt{2}$,
i.e.\ $B  \leq  6$ T in GaAs \cite{Laughlin83:3383}. 
However, in FQHE experimental conditions, the magnetic field is much higher. 
In  \cite{Du93:70}, the energy gaps of FQHE states  related to samples A  and B at 
filling factors $p/(2p\pm1)$ between $\nu=\frac{1}{4}$ and $\nu=\frac{1}{2}$ are
shown to increase linearly with the deviation of $B$ from the  respective  characteristic
values $B_{\frac{1}{2}}^A=9.25$ T, $B_{\frac{1}{4}}^A=18.50$ T
and $B_{\frac{1}{2}}^B=19$ T (where the superscripts refer to the samples). 
The corresponding slopes respectively yield the direct measures $M_A=0.63$,
$M_A=0.93$  and $M_B=0.92$ of the effective electron mass in units of the electron mass $m_e$.  
Indeed, since these masses scale like $\lambda_c^{-1}$ and hence 
like $\sqrt{B}$ for they are determined by electron-electron interaction,
we have $0.63/\sqrt{9.25}=0.207\approx 0.93/\sqrt{18.50}=0.216 \approx  0.92/\sqrt{19}=0.211$. 
Therefore, introducing the parameter $\kappa$ that accounts for the above
experimental results, we have \cite{Du93:70}:
\begin{equation}
\label{eq-MvsSQRTofB}
      \frac{M}{m_e} \sim \kappa\sqrt{B}\quad ;\quad 0.207\leq \kappa\leq 0.216,
\end{equation}
where $B$ is given in Tesla.
\begin{figure}[htbq]
      \includegraphics[width=0.48\textwidth,angle=0,bb=20 20 1200 921,clip]{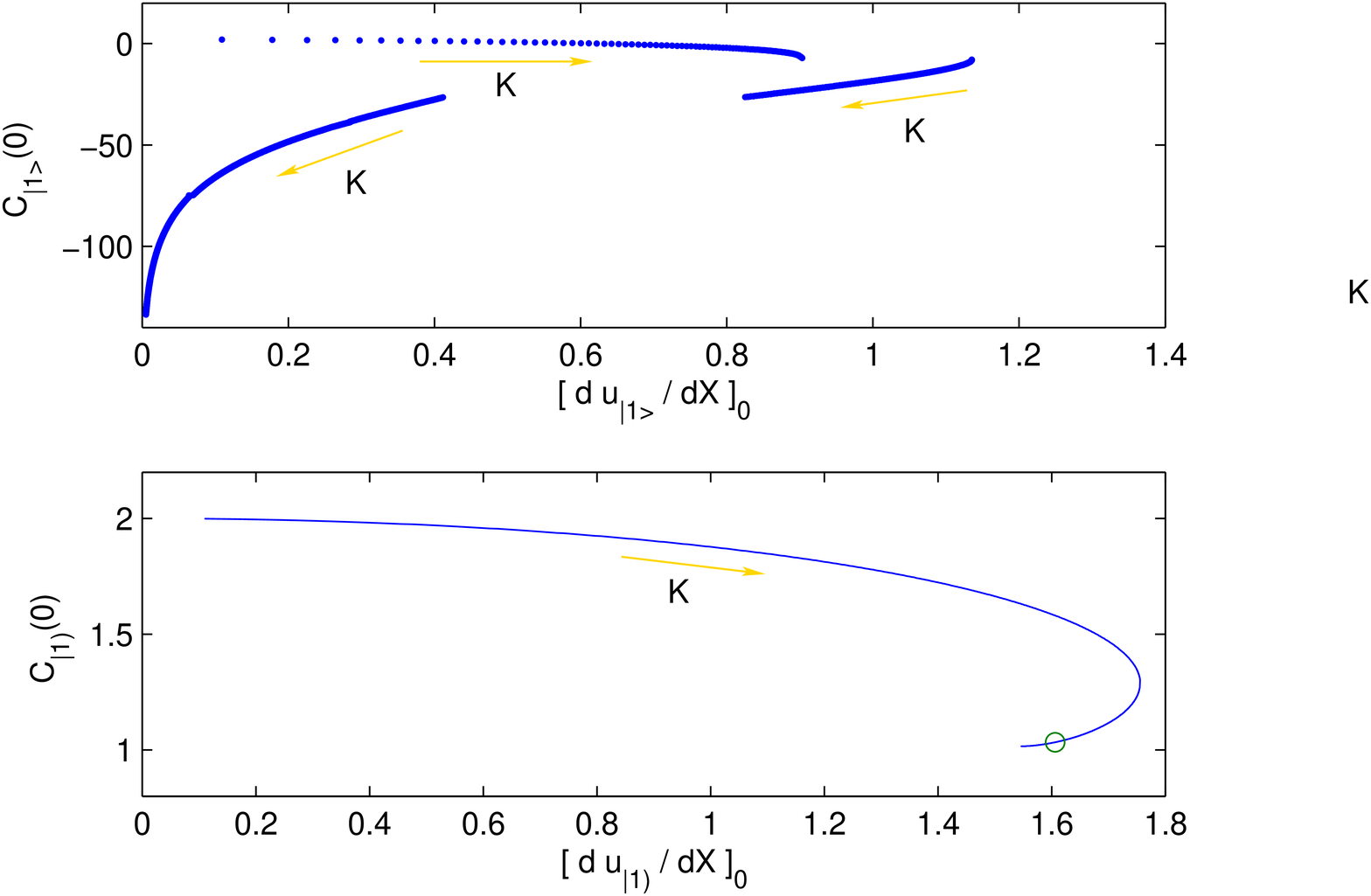}
      \caption{Upper plot: the ``trajectory'' 
      $C_{|1\rangle}(0)$ vs $[du_{|1\rangle}/dX]_0$ for increasing $K$ values whithin the interval $[ 0,15 ]$
      as indicated by the arrows.  It is
      defined by the initial conditions of (\ref{eq-SchroeRBL}),
      or equivalently of (\ref{eq-nlSchroeLAUGH}-\ref{eq-CspLAUGH}).
      Lower plot: the corresponding  ``trajectory''
      defined by the initial conditions of the SP differential system
      (\ref{eq-nlSchroeLAUGH}), (\ref{eq-CspLAUGH}) and (\ref{eq-nlPois}). 
      The circle indicates
       the $\nu=\frac{1}{3}$ FQHE solution defined by (\ref{eq-Kvalue}).
      In both plots, the $K=0$ free-electron case
      is defined by the upper left point $[du/dX]_0 = 0 $ and $C(0)=2$.}
\label{fig1}
\end{figure}

Equation (\ref{eq-SchroeRBL}) is linear and hence dispersive
in its free-particle angular-momentum eigenspace.  Its
stationary solutions  are expected to spread out over more and more eigenstates $|m\rangle$
when the  perturbation defined by $ K\neq 0$ grows. This is best illustrated  by
Fig.\ \ref{fig1} (upper plot).
Starting at  $K= 0$ (no interaction) from $m=1$ lowest-energy and most stable free-electron
vortex state $|1\rangle$ defined by (\ref{eq-SchroeRBL}), it
implicitely displays in terms of increasing $K$ the ``trajectory'' corresponding to 
the solution of (\ref{eq-SchroeRBL}) in its initial-condition phase space.
Let us rewrite (\ref{eq-SchroeRBL}) under the form of  
the following equivalent differential system:
\begin{equation}
\label{eq-nlSchroeLAUGH}
      \Bigl[\nabla_X^2 + C_{|1\rangle} -\frac{1}{X^2} -\frac{X^2}{4} \Bigr] u_{|1\rangle}=0, 
\end{equation}
\begin{equation}
\label{eq-nlPoisLAUGH}
      \nabla_X^2 C_{|1\rangle} = K \delta(X), 
\end{equation}
with 
\begin{equation}
\label{eq-CspLAUGH}
      C_{|1\rangle}(X)=\mu_{|1\rangle}  -{\cal W}_{|1\rangle}(X),
\end{equation}
where $\delta(X)$ is the Dirac function, the radial Laplacian $\nabla_X^2$ 
is 3D in (\ref{eq-nlPoisLAUGH})
while it remains 2D in (\ref{eq-nlSchroeLAUGH}) (this point will be discussed
further below), the eigenvalue  $\mu_{|1\rangle}$ stands for $E _{|1\rangle}+1$ 
due to the Larmor rotation at $m=1$
and ${\cal W}_{|1\rangle}=K/X$ describes the particle-particle interaction potential 
defined by the last term in (\ref{eq-SchroeRBL}). Then the initial-condition
phase space becomes $C_{|1\rangle}(0)$ vs $[du_{|1\rangle}/dX]_0$.
Indeed, we let $u(0)=[dC/dX]_0=0$  
due to, respectively, complete depletion in the vortex trough and radial symmetry 
(no cusp). 
There are clearly two discontinuities in Fig.\ \ref{fig1} (upper plot)   which
describe the ``jumps'' of the initial $m=1$
solution to higher orbital momenta when $K$ grows. In particular, there is a
phase transition (infinite slope) at $[du_{|1\rangle}/dX]_{0}\sim 0.85$
and $C_{|1)}(0)\sim -7$.
Now compare with  Fig.\ \ref{fig1} (lower plot). It displays the trajectory 
of the nonlinear Schr{\"o}dinger-Poisson (SP)   solution 
which is defined from (\ref{eq-nlSchroeLAUGH}-\ref{eq-CspLAUGH})
by adding  the  mean-field source term $u_{|1)}^2$ to Poisson equation (\ref{eq-nlPoisLAUGH}), namely:
\begin{equation}
\label{eq-nlPois}
      \nabla^2 C_{|1)}  =  u_{|1)}^2,
\end{equation}
(we emphasize the eigenstate's  nonlinear nature imposed  by Eq. (\ref{eq-nlPois}) by  using
parentheses instead of kets). 
\begin{figure*}[htbq]
      \includegraphics[width=0.8\textwidth,angle=0,bb=90 260 550 570]{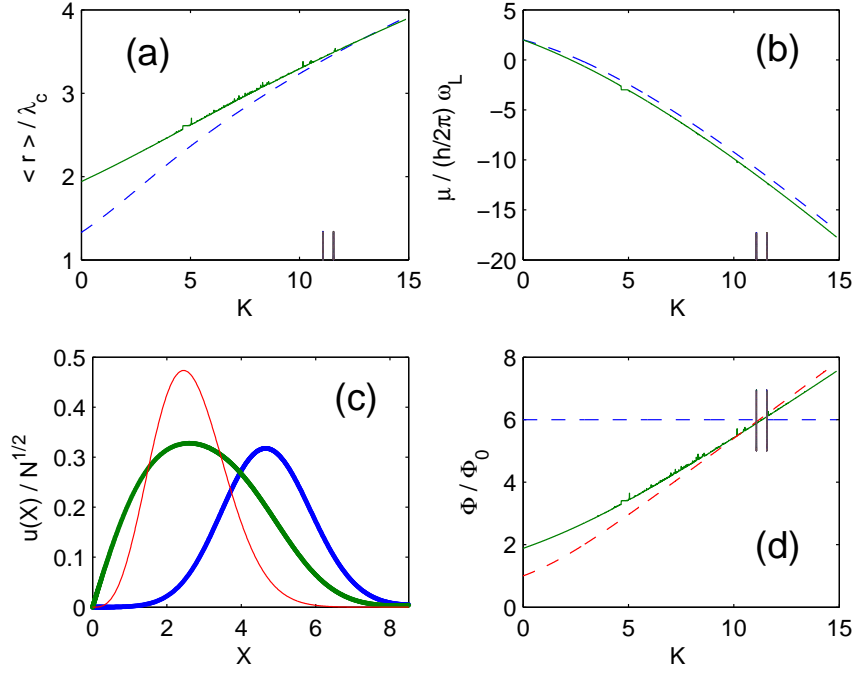}
      \caption{The nonlinear SP model for CF  illustrated by its four  basic properties. The experimental
      range (\ref{eq-Kvalue}) of nonlinearity $K$ is displayed by the two vertical marks. {\bf (a)}:
      the ``nonlinear'' radius ${\bar X}$ defined by (\ref{eq-Xbarre}) (continuous line) compared with  
      ${\bar Y}$ defined by (\ref{eq-Ybarre}) (broken line). {\bf(b)}:
      the ``nonlinear'' eigenvalue $\mu_{|1)}$ defined by  
      (\ref{eq-nlSchroeLAUGH}), (\ref{eq-CspLAUGH}) and (\ref{eq-nlPois}) (continuous line)
      compared with $\mu_{|1\rangle}$ defined by
      (\ref{eq-nlSchroeLAUGH}-\ref{eq-CspLAUGH}) (broken line).
      {\bf(c)}: the FQHE normalized profiles $|1)$ (left: bold green) and $|1\rangle$ (right: bold blue)
      at $K={\cal N}_{|1)}=11.23$ (broken-line profiles in Fig.\ \ref{fig3}), compared with the modulus of 
      Laughlin's $\nu=\frac{1}{3}$ normalized wave function (\ref{eq-jastrow}) (thin red line).
      {\bf(d)}: the ``nonlinear" magnetic flux $\Phi_{|1)}$ given by (\ref{eq-Xbarre}) and (\ref{eq-fluxNL}) in
      continuous green line, compared with $\Phi_{|1\rangle}$ given by (\ref{eq-Ybarre}) and (\ref{eq-fluxNL})
      in broken red line. The horizontal broken blue line at $\Phi/\Phi_0=6$
      refers to (\ref{eq-jastrow}) whose flux does not depend on $K$ and yields $6\Phi_0$ \cite{Laughlin87}.}
\label{fig2}
\end{figure*}
The spectral coherence of the new solution ---i.e.\ the invariance of its angular momentum 
with respect to the increase of $K$--- is obvious: instead of discontinuously spreading out in
the momenta space like in  Fig.\ \ref{fig1} (upper plot), the SP solution
starts spiraling down while keeping its $m=1$ initial
value \cite{Reinisch12:902}.
No phase transition towards higher angular momenta occurs  
for $0\leq K \leq 15$
(e.g.\ in Fig.\ \ref{fig1}, lower plot, at $[du_{|1)}/dX]_{0}\sim 1.75$ and $C_{|1)}(0)\sim 1.3$).

\begin{figure}[htbq]
      \includegraphics[width=0.48\textwidth,angle=0,bb=25 25 530 490]{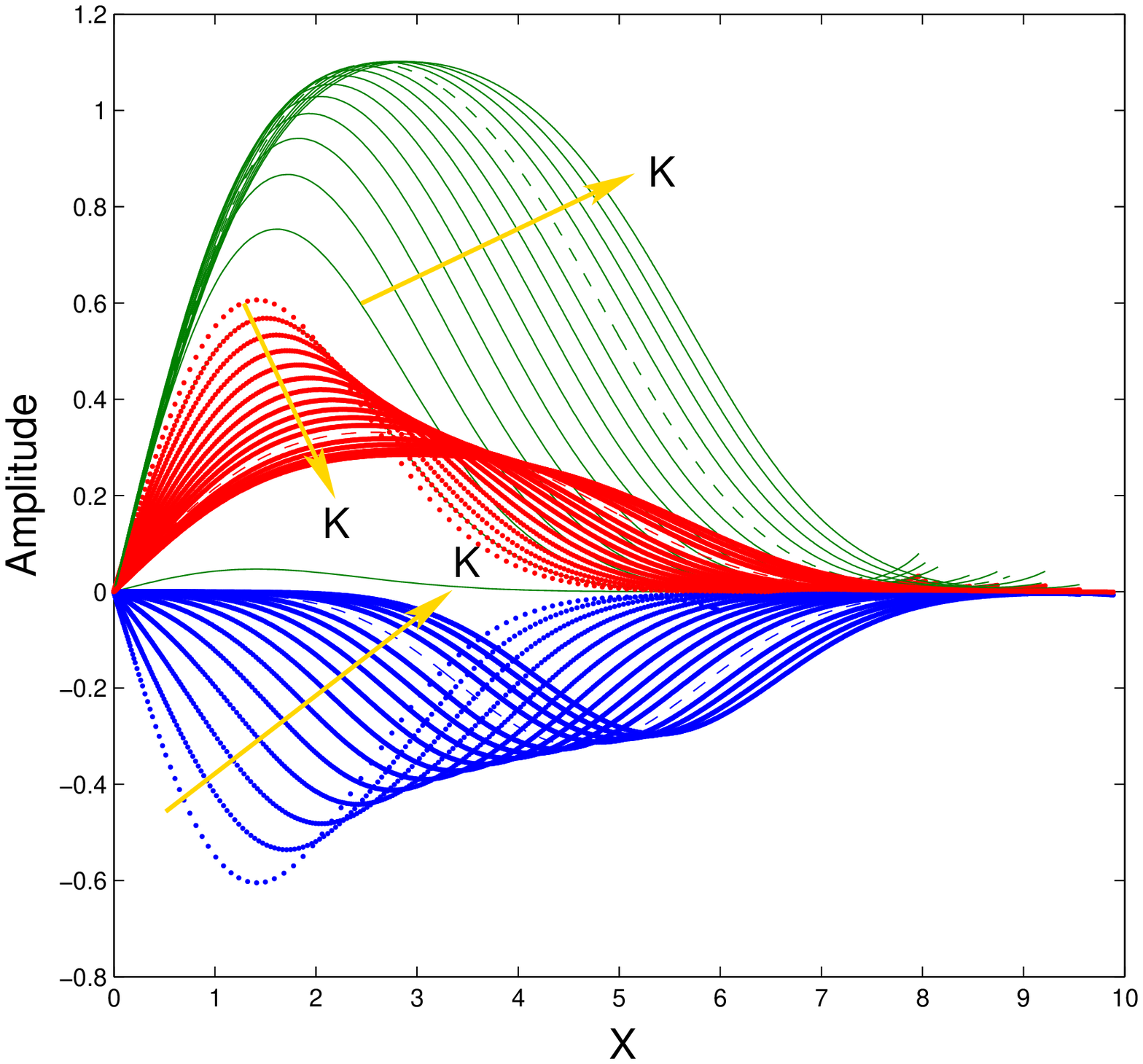}
      \caption{Left: The broadening and flatening of the normalized linear profile
      $-u_{|1\rangle}(X)/\sqrt{{\cal N}_{|1\rangle}}$ corresponding to eigenstate $|1\rangle$ 
      of (\ref{eq-nlSchroeLAUGH})-(\ref{eq-CspLAUGH}) (in bold blue 
      and multiplied by $-1$ for the sake of clarity), compared with the normalized
      nonlinear SP wave function $u_{|1)}(X)/\sqrt{{\cal N}_{|1)}}$ of 
      (\ref{eq-nlSchroeLAUGH}), (\ref{eq-CspLAUGH}) and (\ref{eq-nlPois}) (in bold red) 
      when ${\cal N}_{|1)}=K$ increases by integer steps from $0$ (free-electron solution:
      dotted profiles) to $15$. 
      The SP solution $u_{|1)}(X)$ is displayed in green and increases with $K$.
      The CF profiles defined by ${\cal N}_{|1)}= K=11.23$ are displayed in broken lines. }
\label{fig3}
\end{figure}

This phenomenon resembles the well-known soliton coherence
in hydrodynamics due to the cancellation of the dispersive effects by nonlinearity.
The mathematical tool that explains the stability of the resulting solitary wave
is the so-called nonlinear spectral transform. It provides a theoretical 
link between linear spectral ---and nonlinear dynamical
and/or structural properties of the wave
\cite{Remoissenet99}. This is what our nonlinear transformation
from (\ref{eq-nlPoisLAUGH}) to (\ref{eq-nlPois}) is doing. It introduces
an explicit ab-initio nonlinearity that cancels the angular-momentum
dispersion displayed by Fig.\ \ref{fig1} (incidently, this transformation 
is usually done the opposite way in classical soliton physics: one starts with
the ``real'' nonlinear wave equation and ends up with its formal ``spectral transformed''
linear counterpart \cite{Remoissenet99}).
As shown by Fig.\ \ref{fig2}a, the resulting nonlinear eigenstate $|1)$ defined by the
SP differential system (\ref{eq-nlSchroeLAUGH}), (\ref{eq-CspLAUGH})  and (\ref{eq-nlPois})
yields an average spatial extension which fits with 
the prediction of the linear equation (\ref{eq-SchroeRBL}) provided 
$K$ is chosen in the following FQHE experimental  range 
defined by (\ref{eq-K}-\ref{eq-MvsSQRTofB}):
\begin{equation}
\label{eq-Kvalue}
      11.07\leq K=
      \frac{4\pi^{3/2}M c^2}{\epsilon \Phi_0^{3/2}\sqrt{B}}=\kappa\frac{4\pi^{3/2} 
      m_e c^2}{\epsilon \Phi_0^{3/2}}\leq 11.56.
\end{equation}
This range of relevant $K$ values is indicated  by the circle in Fig.\ \ref{fig1} (lower plot) and by
the two vertical marks in Figs\ \ref{fig2}a,b,d.
Most important for the aim of the present letter, the nonlinear transformation
from (\ref{eq-nlPoisLAUGH}) to (\ref{eq-nlPois})
yields the expected CF properties about
gap stability and flux quantization (see Fig.\ \ref{fig2}b and  2d, respectively)
while the spatial extension of the nonlinear eigenstate $|1)$ also agrees with Laughlin's 
$\nu=\frac{1}{3}$ two-electron normalized wavefunction 
whose modulus $|\Psi_3|$ is derived from \cite{Chakraborty95}:
\begin{equation}
\label{eq-jastrow}
      \Psi_3\propto (z_1-z_2)^{\displaystyle 3}\, e^{\displaystyle  -\frac{(|z_1|^2 + |z_2|^2)}{4\lambda_c}}
      \propto X^{\displaystyle 3} e^{\displaystyle 3i\phi}e^{\displaystyle -\frac{X^2}{4}},
 \end{equation}
up to the unimportant  factor $\exp[\frac{1}{2}({\bar z}/\lambda_c)^2]$ related to the external
degree of freedom. Figure \ref{fig2}c  indeed
displays the FQHE states defined by (\ref{eq-Kvalue}) (broken-line profiles in
Fig.\ \ref{fig3}), namely
$|1)$ (nonlinear: bold green)  and  $|1\rangle$ (linear: bold blue),
as compared to normalized $|\Psi_3|$ (thin red). 

The amplitude of the wavefunction defined by Eqs (\ref{eq-nlSchroeLAUGH}), (\ref{eq-CspLAUGH}) 
and (\ref{eq-nlPois})  reads $\Psi_{|1)}=u_{|1)}\sqrt{M \omega_L/(\pi\hbar {\cal N}_{|1)})}$ 
where \cite{Reinisch12:902}
\begin{equation}
\label{eq-normeDEi}
      {\cal N}_{|1)}=\int_0^{\infty} u_{|1)}^2 XdX,
\end{equation}
in order to achieve  
normalization  according to  $\int_0^{\infty}|| \Psi_{|1)}(x,y)||^2 dx dy = 1$.
Norm (\ref{eq-normeDEi}) is the nonlinear order parameter of our SP model \cite{Reinisch12:902}. 
As it grows, the amplitude and width of  
$u_{|1)}$ increases while its corresponding normalized profile
$u_{|1)}/\sqrt{{\cal N}_{|1)}}$ spreads out: see Fig.\ \ref{fig3}.
Comparing the (last) interaction term of the bracket in (\ref{eq-SchroeRBL})
with the asymptotic solution
\begin{equation}
\label{eq-PoissonASYMPT}
      \lim_{X\rightarrow \infty}{\cal W}_{|1)}(X)={\cal N}_{|1)} G(X)
\end{equation}
of (\ref{eq-CspLAUGH}-\ref{eq-nlPois}) where the 3D Green function defined by (\ref{eq-nlPoisLAUGH}) is
$G(X)=X^{-1}$, we obtain:
\begin{equation}
\label{eq-Nvalue}
      {\cal N}_{|1)}=K.
\end{equation}
Equation (\ref{eq-Nvalue}) operates the link between the two-particle linear 
description (\ref{eq-SchroeRBL}) and the mean-field single-(quasi)particle provided by  
(\ref{eq-nlSchroeLAUGH}), (\ref{eq-CspLAUGH})  and (\ref{eq-nlPois}).
In \cite{Du93:70} the filling factor $\nu=\frac{1}{3}$ lies at the intersection
of two slopes concerning sample A. Taking their average, we obtain
$K \sim 11.23$ which will be our reference value in interval (\ref{eq-Kvalue}). 
Note that it largely exceeds $K=\sqrt{2}$ in \cite{Laughlin83:3383}.
Similarly, ${\cal N}_{|1)}=11.23$ in accordance with (\ref{eq-Nvalue}) yields 
a \underbar{strong} nonlinearity: e.g.\ compare it with $2.53$ in quantum-dot helium  
\cite{Reinisch12:902}. This is quite spectacularly illustrated by  Fig.\ \ref{fig3} 
where the linear solution $|1\rangle$ of (\ref{eq-SchroeRBL})
---or equivalently of system (\ref{eq-nlSchroeLAUGH}-\ref{eq-CspLAUGH})---
is compared with solution $|1)$ of the nonlinear SP   system
(\ref{eq-nlSchroeLAUGH}), (\ref{eq-CspLAUGH})  and (\ref{eq-nlPois}) for increasing $K$ values.
The profiles defined by  $\nu=\frac{1}{3}$  FQHE  value $K={\cal N}_{|1)} =11.23$  are
displayed in broken lines: obviously they are strongly modified with respect to the
free-particle ones (in dotted lines).

Let us now be specific about some technical points used in order to obtain the above results.  
In FQHE, the zeros of the Jastrow-type many-electron wave function proposed by Laughlin for 
odd polynomial degree look like 2D charges which repel  each  other
by logarithmic interaction, yielding a negative value for the
energy $E$ per electron \cite{Chakraborty95}.  Consequently we 
solve the Poisson equation (\ref{eq-nlPois}) in 2D, which indeed ensures that
the corresponding Green function becomes $G(X)=-\log(X)$.
Therefore we obtain
the ``fully 2D'' self-consistent SP differential system (\ref{eq-nlSchroeLAUGH}), 
(\ref{eq-CspLAUGH}) and (\ref{eq-nlPois}) whose eigensolution $u_{|1)}(X)$ is defined by   \ref{eq-Kvalue},
(\ref{eq-normeDEi}) and (\ref{eq-Nvalue}).

The radius of SP's nonlinear state $|1)$ displayed by Fig.\ \ref{fig2}a:
\begin{equation}
\label{eq-Xbarre}
      {\bar X}=\sqrt{\langle {\bar z}^2 \rangle}_{|1)}=
      \frac{1}{\sqrt{2}}[\langle X^2\rangle_{|1)}+\langle X\rangle^2_{|1)}]^{\frac{1}{2}},
\end{equation}
is obtained from  ${\bar z}=\frac{1}{2} (z_1+z_2)$ by quantum-averaging $X$ and $X^2$
in the state $|1)$ (hence the subscripts) since the two 
electrons located at $z_1$ and $z_2$ are both in the same  state $|1)$\cite{Griffiths05}.
On the other hand, $|z_1-z_2|$ is the diameter of the two-electron orbit
defined by the linear internal degree of freedom $|1\rangle$ when assuming that the   
external degree of freedom ${\bar z}$ is frozen in  
its ground state. Therefore its radius:
\begin{equation}
\label{eq-Ybarre}
      {\bar Y}=\frac{1}{2} \langle |z_1-z_2| \rangle_{|1\rangle}= \frac{1}{\sqrt{2}}\langle X \rangle_{|1\rangle}
\end{equation}
(broken line in Fig.\ \ref{fig2}a ) can indeed  be compared with (\ref{eq-Xbarre}) (continuous line). 
Like already emphasized, these two radii coincide at the experimental range
(\ref{eq-Kvalue}) displayed by the couple of adjacent vertical marks in Fig.\ \ref{fig2}a. 

The  nonlinear energy eigenvalue  $\mu_{|1)}$ solution of 
(\ref{eq-nlSchroeLAUGH}), (\ref{eq-CspLAUGH})  and (\ref{eq-nlPois})
(continuous line in Fig.\ \ref{fig2}b ) is obtained from (\ref{eq-CspLAUGH})  
by use of the 2D Green function $G(X)=-\log(X)$ either from the initial condition
$C_0=C_{|1)}(0)$ with
${\cal W}_{|1)}(0)=\int_0^{\infty} G(X) u_{|1)}^2(X) dX$ in agreement with (\ref{eq-nlPois}); 
or at the boundary $C_{|1)}(X\rightarrow \infty) $ by use of 
(\ref{eq-PoissonASYMPT}). The equivalence of these two definitions constitutes 
a test for the relevance of our numerical code: they fit within a relative error of $10^{-7}$.
On tne other hand, the energy eigenvalue $\mu_{|1\rangle}$
corresponding to the solution  $u_{|1\rangle}$ of the linear differential system
(\ref{eq-nlSchroeLAUGH}-\ref{eq-CspLAUGH}) can also be obtained from (\ref{eq-CspLAUGH})  
at the two above limits by simply using the explicit 2D definition 
${\cal W}_{|1\rangle}(X)=-K \log(X)$. It is displayed in Fig.\ \ref{fig2}b by the broken line.
The negative energy gap $\Delta=\mu_{|1)}-\mu_{|1\rangle}$ yields the stability
of $|1)$ when compared with $|1\rangle$.
Though small, it is clearly visible.
We obtain at the experimental FQHE value $K=11.23$:  $\mu_{|1\rangle}=-11.0807$  
$\hbar\omega_L$ $=-0.6977$  $e^2/\epsilon \lambda_c$ and $\mu_{|1)}=-11.7164$  
$\hbar\omega_L$ $=-0.7377$ $e^2/\epsilon \lambda_c$ (cf.\ (\ref{eq-K})).
Therefore $\Delta=-0.04$ $e^2/\epsilon \lambda_c$, to be compared with some
experimental value $\Delta\sim -0.1$ $e^2/\epsilon \lambda_c$ \cite{Du93:70}. 
Moreover, the nonlinear eigenenergy per particle in our $N_e=2$ system is
$E_2=\frac{1}{2}\mu_{|1)}=-0.37$ $e^2/\epsilon \lambda_c$. 
In the $20\leq N_e\leq 144$ interacting electron case
in the disk geometry with the filling factor $\nu=\frac{1}{3}$, 
$E_2\sim -0.39$  $e^2/\epsilon \lambda_c$
by extrapolation to $N_e=2$ of the Monte Carlo evaluation
of the ground-state energy per particle \cite{Morf86:33}.
On the other hand, $E\sim -0.41$  $e^2/\epsilon \lambda_c$ per particle is almost insensitive
to the system size for $4\leq N_e\leq 6$ \cite{Yoshioka83:50}. Therefore our $E_2=-0.37$ $e^2/\epsilon \lambda_c$  
seems quite acceptable in this context. 

Figure \ref{fig2}d displays the fundamental property of the present model, namely,
its $6\Phi_0$ flux quantization for the LLL filling factor $\nu=\frac{1}{3}$  
in the experimental range defined by  \ref{eq-Kvalue}. 
Indeed we have respectively from (\ref{eq-Xbarre}) and (\ref{eq-Ybarre}):
\begin{equation}
\label{eq-fluxNL}
      \frac{\Phi_{|1)}}{\Phi_0}=\frac{\pi B \lambda_c^2}{\Phi_0} \,{\bar X}^2=
      \frac{1}{2}{\bar X}^2=6,
\end{equation}
\begin{equation}
\label{eq-fluxL}
      \frac{\Phi_{|1\rangle}}{\Phi_0}=\frac{\pi B \lambda_c^2}{\Phi_0}\, {\bar Y}^2=
      \frac{1}{4} \langle X \rangle_{|1\rangle}^2=6,
\end{equation}
between the two vertical marks.
The (blue) horizontal broken line at $\Phi/\Phi_0=6$ refers to Laughlin's 
$\nu=\frac{1}{3}$ normalized wavefunction ansatz (\ref{eq-jastrow})
which does not depend on $K$. Its flux
is indeed $6\Phi_0$ \cite{Laughlin87}). 

In conclusion, we stress the self-consistency of  
our FQHE nonlinear model. The spectral coherence of the mean-field SP mode $|1)$ 
slightly lowers the energy per electron with respect to that obtained from 
Schr{\"o}dinger's two-electron internal mode $|1\rangle$. This gap
makes the nonlinear mode $|1)$ energetically favourable  for
the fulfilment of the flux quantization condition (\ref{eq-fluxNL}) than
the linear SP mode $|1\rangle$ for (\ref{eq-fluxL}). 
This property might be considered as the attachment of two flux quanta per electron and
the subsequent transformation of this later into a nonlinear soliton-like  
CF whose 3rd remaining flux quantum makes it behave as a mere quasiparticle in LLL
\underbar{Integral} Quantum Hall Effect.

\begin{acknowledgments}
\noindent GR gratefully acknowledges Lagrange lab's hospitality at the
Observatoire de la Cote d'Azur (Nice, France). The authors acknowledge 
financial support from the Icelandic Research and Instruments Funds,
the Research Fund of the University of Iceland.
\end{acknowledgments}


\begin{thebibliography}{19}
\expandafter\ifx\csname natexlab\endcsname\relax\def\natexlab#1{#1}\fi
\expandafter\ifx\csname bibnamefont\endcsname\relax
  \def\bibnamefont#1{#1}\fi
\expandafter\ifx\csname bibfnamefont\endcsname\relax
  \def\bibfnamefont#1{#1}\fi
\expandafter\ifx\csname citenamefont\endcsname\relax
  \def\citenamefont#1{#1}\fi
\expandafter\ifx\csname url\endcsname\relax
  \def\url#1{\texttt{#1}}\fi
\expandafter\ifx\csname urlprefix\endcsname\relax\def\urlprefix{URL }\fi
\providecommand{\bibinfo}[2]{#2}
\providecommand{\eprint}[2][]{\url{#2}}

\bibitem[{\citenamefont{Read}(1989)}]{Read89:62}
\bibinfo{author}{\bibfnamefont{N.}~\bibnamefont{Read}}, \bibinfo{journal}{Phys
  Rev.Lett} \textbf{\bibinfo{volume}{62}}, \bibinfo{pages}{86}
  (\bibinfo{year}{1989}).

\bibitem[{\citenamefont{Jain}(1989)}]{Jain89:199}
\bibinfo{author}{\bibfnamefont{J.~K.} \bibnamefont{Jain}},
  \bibinfo{journal}{Phys Rev.Lett.} \textbf{\bibinfo{volume}{63}},
  \bibinfo{pages}{199} (\bibinfo{year}{1989}).

\bibitem[{\citenamefont{Stormer}(1999)}]{Stormer99:875}
\bibinfo{author}{\bibfnamefont{H.~L.} \bibnamefont{Stormer}},
  \bibinfo{journal}{Rev. Mod. Phys.} \textbf{\bibinfo{volume}{71}},
  \bibinfo{pages}{875} (\bibinfo{year}{1999}).

\bibitem[{\citenamefont{Laughlin}(1983{\natexlab{a}})}]{Laughlin83:1395}
\bibinfo{author}{\bibfnamefont{R.~B.} \bibnamefont{Laughlin}},
  \bibinfo{journal}{Phys Rev.Lett.} \textbf{\bibinfo{volume}{50}},
  \bibinfo{pages}{1395} (\bibinfo{year}{1983}{\natexlab{a}}).

\bibitem[{\citenamefont{Laughlin}(1983{\natexlab{b}})}]{Laughlin83:3383}
\bibinfo{author}{\bibfnamefont{R.~B.} \bibnamefont{Laughlin}},
  \bibinfo{journal}{Phys. Rev. B} \textbf{\bibinfo{volume}{27}},
  \bibinfo{pages}{3383} (\bibinfo{year}{1983}{\natexlab{b}}).

\bibitem[{\citenamefont{Girvin and MacDonald}(1987)}]{Girvin87:1252}
\bibinfo{author}{\bibfnamefont{S.~M.} \bibnamefont{Girvin}} \bibnamefont{and}
  \bibinfo{author}{\bibfnamefont{A.~H.} \bibnamefont{MacDonald}},
  \bibinfo{journal}{Phys Rev.Lett.} \textbf{\bibinfo{volume}{58}},
  \bibinfo{pages}{1252} (\bibinfo{year}{1987}).

\bibitem[{\citenamefont{Zhang et~al.}(1989)\citenamefont{Zhang, Hannsson, and
  Kivelson}}]{Zhang89:62}
\bibinfo{author}{\bibfnamefont{S.~C.} \bibnamefont{Zhang}},
  \bibinfo{author}{\bibfnamefont{T.~H.} \bibnamefont{Hannsson}},
  \bibnamefont{and} \bibinfo{author}{\bibfnamefont{S.}~\bibnamefont{Kivelson}},
  \bibinfo{journal}{Phys Rev.Lett.} \textbf{\bibinfo{volume}{62}},
  \bibinfo{pages}{82} (\bibinfo{year}{1989}).

\bibitem[{\citenamefont{Rajaraman and Sondhi}(1996)}]{Rajaraman96:793}
\bibinfo{author}{\bibfnamefont{R.}~\bibnamefont{Rajaraman}} \bibnamefont{and}
  \bibinfo{author}{\bibfnamefont{S.~L.} \bibnamefont{Sondhi}},
  \bibinfo{journal}{Int. J. Mod. Phys. B} \textbf{\bibinfo{volume}{10}},
  \bibinfo{pages}{793} (\bibinfo{year}{1996}).

\bibitem[{\citenamefont{Halperin et~al.}(1993)\citenamefont{Halperin, Lee, and
  Read}}]{Halperin93:47}
\bibinfo{author}{\bibfnamefont{B.}~\bibnamefont{Halperin}},
  \bibinfo{author}{\bibfnamefont{P.}~\bibnamefont{Lee}}, \bibnamefont{and}
  \bibinfo{author}{\bibfnamefont{N.}~\bibnamefont{Read}},
  \bibinfo{journal}{Phys Rev.B} \textbf{\bibinfo{volume}{47}},
  \bibinfo{pages}{7312} (\bibinfo{year}{1993}).

\bibitem[{\citenamefont{Rezayi and Read}(1994)}]{Rezayi94:72}
\bibinfo{author}{\bibfnamefont{E.~H.} \bibnamefont{Rezayi}} \bibnamefont{and}
  \bibinfo{author}{\bibfnamefont{N.}~\bibnamefont{Read}},
  \bibinfo{journal}{Phys Rev.Lett} \textbf{\bibinfo{volume}{72}},
  \bibinfo{pages}{900} (\bibinfo{year}{1994}).

\bibitem[{\citenamefont{Jacak et~al.}(2012)\citenamefont{Jacak, Gonczarek,
  Jacak, and Jozwiak}}]{Jacak12:26}
\bibinfo{author}{\bibfnamefont{J.}~\bibnamefont{Jacak}},
  \bibinfo{author}{\bibfnamefont{R.}~\bibnamefont{Gonczarek}},
  \bibinfo{author}{\bibfnamefont{L.}~\bibnamefont{Jacak}}, \bibnamefont{and}
  \bibinfo{author}{\bibfnamefont{I.}~\bibnamefont{Jozwiak}},
  \bibinfo{journal}{Int. J. Mod. Phys. B} \textbf{\bibinfo{volume}{26}},
  \bibinfo{pages}{1230011} (\bibinfo{year}{2012}).

\bibitem[{\citenamefont{Du et~al.}(1993)\citenamefont{Du, Stormer, Tsui,
  Pfeiffer, and West}}]{Du93:70}
\bibinfo{author}{\bibfnamefont{R.~R.} \bibnamefont{Du}},
  \bibinfo{author}{\bibfnamefont{H.~L.} \bibnamefont{Stormer}},
  \bibinfo{author}{\bibfnamefont{D.~C.} \bibnamefont{Tsui}},
  \bibinfo{author}{\bibfnamefont{L.~N.} \bibnamefont{Pfeiffer}},
  \bibnamefont{and} \bibinfo{author}{\bibfnamefont{K.~W.} \bibnamefont{West}},
  \bibinfo{journal}{Phys Rev.Lett.} \textbf{\bibinfo{volume}{70}},
  \bibinfo{pages}{2944} (\bibinfo{year}{1993}).

\bibitem[{\citenamefont{Laughlin}(1987)}]{Laughlin87}
\bibinfo{author}{\bibfnamefont{R.~B.} \bibnamefont{Laughlin}},
  \emph{\bibinfo{title}{The Quantum Hall Effect}}
  (\bibinfo{publisher}{Springer}, \bibinfo{year}{1987}).

\bibitem[{\citenamefont{Reinisch and Gudmundsson}(2012)}]{Reinisch12:902}
\bibinfo{author}{\bibfnamefont{G.}~\bibnamefont{Reinisch}} \bibnamefont{and}
  \bibinfo{author}{\bibfnamefont{V.}~\bibnamefont{Gudmundsson}},
  \bibinfo{journal}{Physica D} \textbf{\bibinfo{volume}{241}},
  \bibinfo{pages}{902} (\bibinfo{year}{2012}).

\bibitem[{\citenamefont{Remoissenet}(1999)}]{Remoissenet99}
\bibinfo{author}{\bibfnamefont{M.}~\bibnamefont{Remoissenet}},
  \emph{\bibinfo{title}{Waves called solitons: concepts and experiments}}
  (\bibinfo{publisher}{Springer}, \bibinfo{year}{1999}).

\bibitem[{\citenamefont{Chakraborty and Pietil\"ainen}(1995)}]{Chakraborty95}
\bibinfo{author}{\bibfnamefont{T.}~\bibnamefont{Chakraborty}} \bibnamefont{and}
  \bibinfo{author}{\bibfnamefont{P.}~\bibnamefont{Pietil\"ainen}},
  \emph{\bibinfo{title}{The quantum Hall effects: integral and fractional}}
  (\bibinfo{publisher}{Springer}, \bibinfo{year}{1995}).

\bibitem[{\citenamefont{Griffiths}(2005)}]{Griffiths05}
\bibinfo{author}{\bibfnamefont{D.~J.} \bibnamefont{Griffiths}},
  \emph{\bibinfo{title}{Introduction to quantum mechanics (2nd Ed.}}
  (\bibinfo{publisher}{Pearson Prentice Hall}, \bibinfo{year}{2005}).

\bibitem[{\citenamefont{Morf and Halperin}(1986)}]{Morf86:33}
\bibinfo{author}{\bibfnamefont{R.}~\bibnamefont{Morf}} \bibnamefont{and}
  \bibinfo{author}{\bibfnamefont{B.~I.} \bibnamefont{Halperin}},
  \bibinfo{journal}{Phys Rev. B} \textbf{\bibinfo{volume}{33}},
  \bibinfo{pages}{2221} (\bibinfo{year}{1986}).

\bibitem[{\citenamefont{Yoshioka et~al.}(1983)\citenamefont{Yoshioka, Halperin,
  and Lee}}]{Yoshioka83:50}
\bibinfo{author}{\bibfnamefont{D.}~\bibnamefont{Yoshioka}},
  \bibinfo{author}{\bibfnamefont{B.~I.} \bibnamefont{Halperin}},
  \bibnamefont{and} \bibinfo{author}{\bibfnamefont{P.~A.} \bibnamefont{Lee}},
  \bibinfo{journal}{Phys Rev.Lett.} \textbf{\bibinfo{volume}{50}},
  \bibinfo{pages}{1219} (\bibinfo{year}{1983}).

\end{thebibliography}
\end{document}